# Indium-Tin-Oxide Transistors with One Nanometer Thick Channel and Ferroelectric Gating

Mengwei Si[1], Joseph Andler[2], Xiao Lyu[1], Chang Niu[1], Suman Datta[3], Rakesh Agrawal[4], Peide D. Ye[1,*]

[1]*School of Electrical and Computer Engineering and Birck Nanotechnology Center, Purdue University, West Lafayette, IN 47907, United States*

[2]*School of Materials Engineering, Purdue University, West Lafayette, IN 47907, United States*

[3]*Department of Electrical Engineering, University of Notre Dame, Notre Dame, IN 46556, United States*

[4]*Davidson School of Chemical Engineering, Purdue University, West Lafayette, IN 47907, United States*

* Address correspondence to: yep@purdue.edu (P.D.Y.)

ABSTRACT

In this work, we demonstrate high performance indium-tin-oxide (ITO) transistors with the channel thickness down to 1 nm and ferroelectric $Hf_{0.5}Zr_{0.5}O_2$ as gate dielectric. On-current of 0.243 A/mm is achieved on sub-micron gate-length ITO transistors with a channel thickness of 1 nm, while it increases to as high as 1.06 A/mm when the channel thickness increases to 2 nm. A raised source/drain structure with a thickness of 10 nm is employed, contributing to a low contact resistance of 0.15 Ω·mm and a low contact resistivity of $1.1 \times 10^{-7}$ Ω·cm$^2$. The ITO transistor with a recessed channel and ferroelectric gating demonstrates several advantages over 2D semiconductor transistors and other thin film transistors, including large-area wafer-size nanometer thin film formation, low contact resistance and contact resistivity, atomic thin channel being immunity to short channel effects, large gate modulation of high carrier density by ferroelectric gating, high-quality gate dielectric and passivation formation, and a large bandgap for the low-power back-end-of-line (BEOL) CMOS application.

The complementary metal-oxide-semiconductor (CMOS) scaling is the driven force of the progress in microelectronic technology. Transistors with ultrathin semiconductor channel are well-known to enhance the electrostatic gate control and the immunity to short channel effects, and to further shrink the size of transistors in the integrate circuit. Therefore, 2-dimensional (2D) van der Waals materials, as atomically thin semiconductors, attract wide attentions as channel materials for the next-generation microelectronic device over the past decade.[1–11] However, there are three major limitations for 2D materials that limit the application of 2D materials as transistor channels in CMOS integrate circuits. The first limit is the high contact resistance in 2D transistors in general due to the existing Schottky barriers at metal/2D semiconductor interfaces and the lack of proper doping techniques in 2D materials.[3,4,7] The 2D semiconductor underneath source/drain contact metals is also atomic thin, which limits the current flow to the channel, resulting in a high access resistance. The second challenge is to grow large-area or wafer-scale single crystal 2D semiconductor compatible for CMOS integration.[12] The third drawback is that it is fundamentally difficult to form high-quality gate dielectrics with low interface trap density on top of 2D materials, because there is no dangling bond on 2D surface and the conventional atomic layer deposition cannot be chemically initiated.[11]

Thin-film transistors (TFTs) with oxide semiconductor channel, such as Indium-Gallium-Zinc-Oxide (IGZO), are widely studied for display applications. But there is very rare study to apply thin-film transistor technology for BEOL CMOS digital applications. Indium-Tin-Oxide (ITO) is a *n*-type degenerated semiconductor with an optical bandgap of 3.5-4.3 eV and frequently used as a transparent "metal" layer due to its very high electron doping introduced by Sn. The typical electron density of ITO as a transparent conductor is about $10^{20}$-$10^{21}$ /cm$^3$, which is usually too high as a transistor channel. ITO transistors were reported by forming a more semiconducting type channel with an electron density of ~$10^{19}$ /cm$^3$ and good current on/off ratio as a switch.[13–17] Ferroelectric gating was also reported to enhance the gate controllability.[13] Li *et al.* recently reported a high-performance ITO transistor with channel thickness down to 4 nm. On-current ($I_{ON}$) exceeding 1 A/mm was achieved on a device with 10 nm thick ITO channel at a channel length of 200 nm, demonstrating that ITO TFT can be a promising candidate for low-power high-performance device application.[16] Further reduction of ITO channel thickness is highly demanded to further improve the immunity to short channel effects for transistor scaling. Beyond ITO, scaled devices on W-doped $In_2O_3$ (IWO) or IGZO are also being investigated now.[18,19] More importantly, TFT with oxide semiconductor as channel recently attract revived interest since it can be applied in BEOL compatible transistors for monolithic 3D integration.[20]

In this work, we report ITO transistors with 1-nm and 2-nm thick channel and ferroelectric (FE) hafnium zirconium oxide ($HfZrO_2$ or HZO) as gate insulator. Highly doped ITO with 3D carrier density ($n_{3D}$) of $1.7 \times 10^{20}$ /cm$^3$ is employed, which enables the channel thickness ($T_{ch}$) scaling down to 1 nm. The high polarization charge density in FE HZO[21–25] enhances the gate controllability so that the high carrier density can be fully depleted. A raised source/drain structure is applied so that low contact resistance ($R_c$) of 0.15 Ω·mm and low contact resistivity ($\rho_c$) of $1.1 \times 10^{-7}$ Ω·cm$^2$ are achieved. High $I_{ON}$ of 1.06 A/mm is achieved on ITO transistor with $T_{ch}$=2 nm and 0.243 A/mm with $T_{ch}$=1 nm with sub-micron channel length. Therefore, high performance ITO transistors with low contact resistance and ultrathin channel are obtained simultaneously, which overcomes fundamental challenges of 2D semiconductors. These results suggest ITO as a promising channel material for BEOL CMOS application.

RESULTS AND DISCUSSION

Fig. 1(a) illustrates the schematic diagram of an ITO transistor with recessed 1-nm thick channel and ferroelectric gating. The gate stack includes heavily boron-doped silicon (p+ Si, resistivity < 0.005 Ω·cm) as gate electrode and 20 nm FE HZO/3 nm Al$_2$O$_3$ as gate insulator. 80 nm Ni is used for source/drain electrodes. The thickness of ITO underneath the source/drain electrodes is 10 nm while the thickness of ITO channel is 1 nm or 2 nm. Fig. 1(b) shows photo images of fabricated ITO transistor, Hall bar and transmission line model (TLM) structures, capturing the ITO channel, Ni electrodes and SiO$_2$ for test pads isolation.

Fig. 1(c) shows the R$_{xy}$ *versus* B field in Hall measurement of 10 nm ITO with floating gate at room temperature. From the Hall measurement, bulk doping concentration n$_{3D}$ of $1.7 \times 10^{20}$ /cm$^3$ and Hall mobility (μ$_{Hall}$) of 16.7 cm$^2$/V·s are determined. Without considering the surface states, we can estimate n$_{2D}$ of $1.7 \times 10^{13}$ /cm$^2$ for a 1nm-channel at zero gate bias, which is about the right carrier density range to be modulated and controlled by an electrostatic gate, suggesting recess channel can be used to enhance the on/off ratio. Fig. 1(d) shows resistance (R) *versus* length (L) characteristics in TLM measurement of 10 nm ITO with Ni contacts. R$_c$ of 0.15 Ω·mm, sheet resistance (R$_{sh}$) of 2114 Ω/□ and transfer length (L$_T$) of 0.07 μm are extracted by linear fitting of R with respect to L. Contact resistivity (ρ$_c$) of $1.1 \times 10^{-7}$ Ω·cm$^2$ are achieved according to $L_T = \sqrt{\rho_c/R_{sh}}$. Due to near metallic ITO characteristic, obtained R$_c$ and ρ$_c$ are much better than those typically obtained from 2D van der Waals materials.[4,7,10]

Fig. 2(a) shows a transmission electron microscopy (TEM) image of Ni/Al$_2$O$_3$/HZO/Si stack, capturing the amorphous Al$_2$O$_3$ and polycrystalline HZO. Fig. 2(b) shows the X-ray diffraction (XRD) spectrum of FE HZO, confirming the HZO crystal contains orthorhombic phase, which leads to the ferroelectricity of HZO. Fig. 2(c) shows the P-V measurement of 20 nm HZO/3 nm Al$_2$O$_3$ capacitor with 10 nm ITO as top electrode and p+ Si as the bottom electrode, where the voltage is applied to the p+ Si electrode. The high polarization and ferroelectric hysteresis loop confirm the ferroelectricity of this structure. The corresponding C-V measurement at 1 kHz on the same device is shown in Fig. S1 in supporting information, showing a typical ferroelectric C-V hysteresis loop. Capacitance at negative voltage is lower than that at positive voltage because of the depletion of ITO as a degenerated semiconductor. Note that maximum polarization over 20 μC/cm$^2$ corresponds to 2D electron density (n$_{2D}$) over 10$^{14}$ /cm$^2$. The P-V measurement gives two clear indications except for the confirmation of ferroelectricity. The first is n$_{2D}$ in 10-nm ITO is higher than 10$^{14}$ /cm$^2$, which is consistent with Hall measurement in Fig. 1(c), suggesting a recess channel is necessary for sufficient gate control. ITO transistor with T$_{ch}$ of 10 nm cannot be switched off by both conventional gating and ferroelectric gating, as shown in Fig. S3 in supporting information. The second indication is HZO/Al$_2$O$_3$/ITO oxide/oxide interface has a relatively low interface trap density compared to the FE polarization density. So, it could achieve a gate control of n$_{2D}$ over 10$^{14}$/cm$^2$, similar to ion-liquid gating. Such gate control by ferroelectric polarization plays an important role to realize the high-performance ITO transistor in this work.

Fig. 3(a) and 3(b) shows the I$_D$-V$_{GS}$ and I$_D$-V$_{DS}$ characteristics of an ITO transistor with channel length (L$_{ch}$) of 0.6 μm and T$_{ch}$ of 2 nm, exhibiting I$_{ON}$ of 1.06 A/mm and on/off ratio over 6 orders of magnitude. I$_D$ in off-state is the result of gate leakage current, as shown in Fig. S12, suggesting optimizing the gate stack can further improve the on/off ratio. The inset of Fig. 3(a) is the AFM measurement of the 2-nm thick ITO channel. Fig. 3(c) and 3(d) shows the I$_D$-V$_{GS}$ and I$_D$-V$_{DS}$ characteristics of an ITO transistor with L$_{ch}$ of 0.8 μm and T$_{ch}$ of 1 nm, exhibiting I$_{ON}$ of 0.243 A/mm. The inset of Fig. 3(c) is the AFM measurement of the 1-nm thick

ITO channel. $I_{ON}$ of ITO transistors with channel thickness at 1 nm to 2 nm range are significantly higher than most of the reported 2D transistors with reasonable on/off ratio.[1-11] The hysteresis in $I_D$-$V_{GS}$ curve in Fig. 3(a) and 3(c) are the result of ferroelectric polarization and the minor hysteresis loop. To further understand this phenomenon, an ITO transistor with $L_{ch}$ of 3 μm and $T_{ch}$ of 2 nm is measured by applying different voltage sweep ranges. As shown in Fig. S6 in supporting information, the $I_D$-$V_{GS}$ characteristics have a smaller hysteresis at reduced voltage sweep range (from 15 V to 4 V). The $I_D$-$V_{GS}$ characteristics of ITO transistors with $T_{ch}$ of 1 nm and 2 nm and $L_{ch}$ from 40 μm down to 1 μm are summarized in Fig. S7 and Fig. S8 in supporting information, showing similar characteristics to short channel devices except for channel length dependent on-current. Fig. 3(e) shows $I_{ON}$ *versus* 1/$L_{ch}$ scaling metrics of ITO transistor with $T_{ch}$ of 1 nm and 2 nm at $V_{DS}$=4 V, suggesting channel length scaling can bring further performance benefit. The $I_{ON}$ *versus* 1/$L_{ch}$ scaling metrics follows a linear trend until 1 μm. Considering the geometry screen length of ITO transistors[16] ($\lambda$=3.3 nm for 2-nm ITO and $\lambda$=2.4 nm for 1-nm ITO), the deviation from 1/$L_{ch}$ scaling at sub-micron channel is likely to be the result of mobility degradation by self-heating effect, instead of short channel effects. Source/drain series resistance ($R_{SD}$) are extracted to be 0.14 Ω·mm and 0.15 Ω·mm for ITO transistors with $T_{ch}$ of 1 nm and 2 nm, by linear fitting of on-resistance *versus* channel length at different $V_{GS}$, as shown in Fig. S9 in supporting information, which is even smaller than the value obtained from TLM measurements in Fig. 1(d).

Fig. 4 shows the effective mobility ($\mu_{eff}$) of 1-nm and 2-nm ITO extracted from drain conductance ($g_d$) in $I_D$-$V_{DS}$ characteristics, where $\mu_{eff} = \frac{g_d L_{ch}}{W_{ch} Q_n}$ and $Q_n$ is the channel mobile charge density and $W_{ch}$ is the channel width. $\mu_{eff}$ of 26.0 cm$^2$/V·s and 6.1 cm$^2$/V·s are achieved for 2-nm and 1-nm thick ITO, respectively. The mobility degradation with decreasing channel thickness is attributed to the increasing surface scattering, as also shown in the surface roughness study in Fig. S2. Mobility could be improved by atomic layer deposition (ALD) surface passivation or reducing the surface roughness. The field-effect mobility ($\mu_{FE}$) are extracted from transconductance ($g_m$) using $\mu_{FE} = \frac{g_m L_{ch}}{W_{ch} C_{ox} V_{DS}}$. $\mu_{FE}$ of 27.0 cm$^2$/V·s and 6.5 cm$^2$/V·s are achieved for 2-nm and 1-nm thick ITO, which are consistent with effective mobilities. Carrier density can be estimated according to $I_D = n_{2D} q \mu E$, where $n_{2D}$ is the 2D carrier density, q is the elementary charge, µ is the mobility, E is the source to drain electric field. According to this equation, the carrier density in 1-nm and 2-nm ITO channel can be calculated as shown in Fig. S10. Carrier density of 1-nm and 2-nm ITO are similar, with maximum $n_{2D}$ about $0.8 \times 10^{14}$ /cm$^2$. Such higher channel carrier density oxide/oxide interface far beyond polar semiconductor interface and the enhanced modulation of carrier density with high on/off ratio is because of the strong ferroelectric polarization switching.

The recess channel and ferroelectric gating device structure is critical to realize high performance and ultrathin body ITO transistor with several advantages compared to 2D semiconductors and low doped ITO transistors: (i) Large-area wafer-size nanometer thin ITO films can be formed by conventional sputtering technique and the process temperature fulfills the BEOL requirement of lower than 350 ºC. (ii) Thick and heavily doped ITO under source/drain electrodes contributes to the low contact resistivity and contact resistance. (iii) Heavily doped ITO enables channel thickness scaling down to 1 nm, comparable to atomic-scale single-layer or bi-layer 2D van der Waals semiconductor channels. (iv) 1-nm thick channel offers excellent immunity to short channel effects which provides a clear route to further scale down the device down to sub-10nm region with much higher device performance. (v) The high polarization density in ferroelectric gate insulator provides sufficient gate modulation of the high carrier density in ITO channel. (vi) ALD passivation and dielectric can be applied to ITO channel easily because it is a 3D material with dangling bond on the surface being different from 2D van der Waals surface. (vii) ITO offers a semiconductor bandgap as high as 3.5-4.3 eV with the potential for power electronics application. These advantages and the demonstrated high performance transistors suggest ITO is a promising oxide channel material for BEOL CMOS applications.

CONCLUSION

In conclusion, high performance ITO transistors with 1-nm and 2-nm thick channel and ferroelectric gating are demonstrated. High $I_{ON}$ of 1.06 A/mm and 0.243 A/mm is achieved on ITO transistors with $T_{ch}$=2 nm and 1 nm, respectively. A raised source/and drain structure is employed so that low contact resistance of 0.15 Ω·mm and low contact resistivity of $1.1 \times 10^{-7}$ Ω·cm$^2$ are achieved. Low contact resistance and ultrathin channel are obtained simultaneously, which overcomes the fundamental limitations of 2D semiconductors. The advantages of ITO transistor suggest ITO as a promising oxide channel material for BEOL CMOS applications.

METHODS

**Device Fabrication.** The device fabrication process started with ALD of 20 nm HZO and 3 nm Al$_2$O$_3$ on p+ Si substrate. The ALD HZO and Al$_2$O$_3$ were deposited at 200 °C, using [(CH$_3$)$_2$N]$_4$Hf (TDMAHf), [(CH$_3$)$_2$N]$_4$Zr (TDMAZr), (CH$_3$)$_3$Al (TMA) and H$_2$O as the Hf, Zr, Al, and O precursors. The Hf$_{1-x}$Zr$_x$O$_2$ film with x=0.5 was achieved by controlling HfO$_2$:ZrO$_2$ cycle ratio to be 1:1. After the deposition of HZO/Al$_2$O$_3$ stack, the samples were annealed at 500 °C in N$_2$ environment for 1 min by rapid thermal annealing. 10 nm ITO was deposited by RF sputtering from a 15 cm diameter target with a composition of 90 wt% $In_2O_3$ and 10 wt% $SnO_2$ and with a target-to-substrate distance of approximately 20 cm. The sputtering power was 650 W in an argon ambient with a working pressure of $2\times10^{-3}$ torr after achieving a base pressure of at least $2\times10^{-6}$ torr. A four-minute pre-sputter was completed for surface contaminant removal and the process was completed without supplemental heating. Device isolation was then performed by wet etching of ITO using hydrochloric acid solution (HCl, 20%). 50 nm $SiO_2$ was then deposited by e-beam evaporation for the isolation of source/drain pads, which can effectively reduce the leakage current and parasitic capacitance introduced by non-ideal test pads. 80 nm Ni was then deposited as source/drain electrodes. Channel recess was done by wet etching using diluted HCl solution (3.4%), so that the etch rate of ITO can be accurately controlled at nanometer scale, where the etch rate is about 1 nm/s, as shown in Fig. S1 in supporting information.

**Device Characterization.** The thickness of the ITO was measured using a Veeco Dimension 3100 atomic force microscope (AFM) system. Electrical characterization was carried out with a Keysight B1500 system with a Cascade Summit probe station.

## ASSOCIATED CONTENT

**Supporting Information**

The supporting information is available free of charge on the ACS Publication website. Additional details for wet etching process and surface roughness, C-V measurement of the gate stack, I-V characteristics of ITO transistors without channel recess and with dielectric gate insulator, minor loops in ITO transistor with ferroelectric gating, channel length and thickness dependent I-V characteristics, series resistance extraction, carrier density of ITO channel, analysis on devices varations are in the supporting information.

The authors declare no competing financial interest.

## AUTHOR INFORMATION

**Corresponding Author**

*E-mail: yep@purdue.edu

**Author Contributions**

P.D.Y. conceived the idea and supervised experiments. J.A. and R.A. did the ITO film sputtering. X.L. did the ALD deposition of HZO and Al$_2$O$_3$. M.S. performed the device fabrication, electrical measurement and data analysis. C.N. and M.S. conducted the Hall measurement. S.D. provided critical technical inputs on the experiments. M.S. and P.D.Y. co-wrote the manuscript and all authors commented on it.

## ACKNOWLEDGEMENTS

The authors would like to thank C.-J. Su and C.-T. Wu for support on TEM imaging and M. A. Capano for support on XRD measurement. The work was supported by SRC nCORE IMPACT Center.

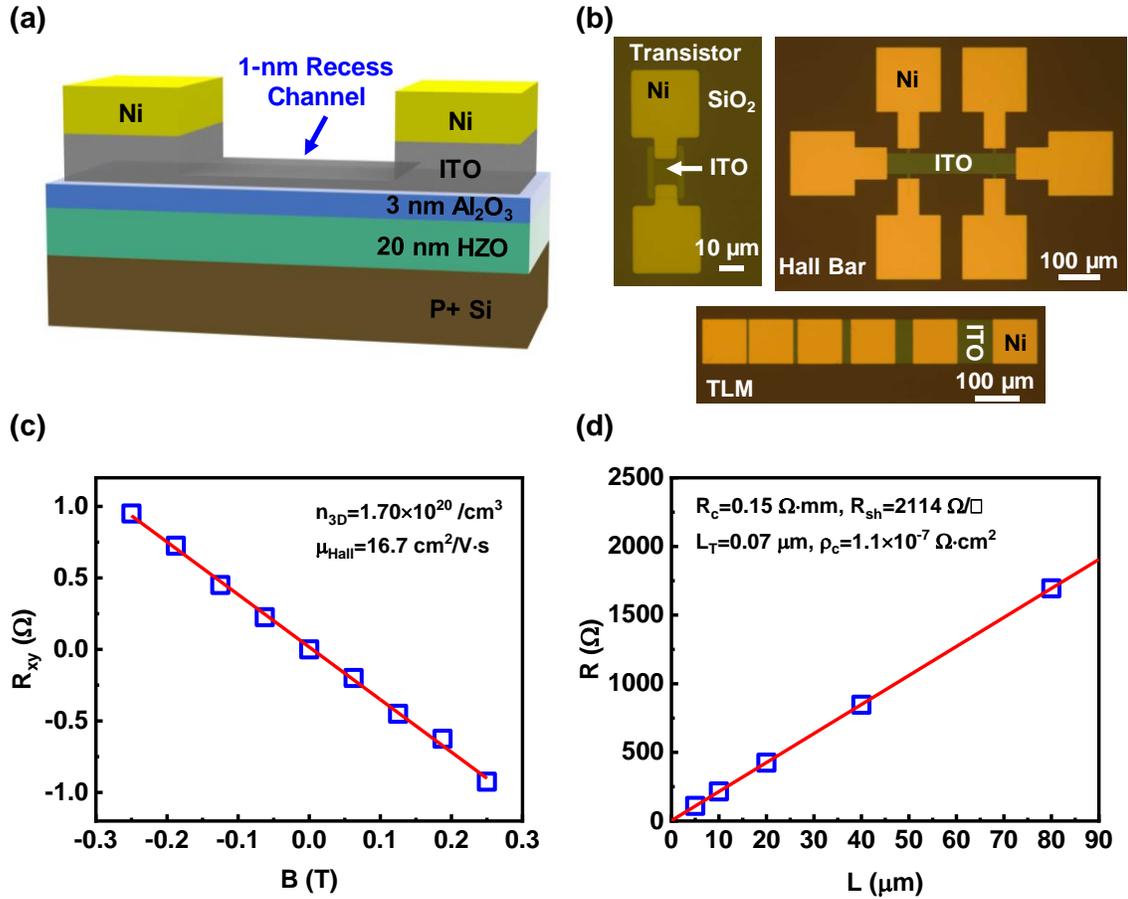

**Figure 1** (a) Schematic diagram of an ITO transistor with 1-nm thick recess channel. The gate stack includes p+ Si as gate electrode and 20 nm ferroelectric HZO/3 nm $Al_2O_3$ as gate insulator. 80 nm Ni is used for source/drain electrodes. The thickness of ITO underneath the source/drain electrodes is 10 nm while the thickness of ITO channel is 1 nm or 2 nm, controlled by wet etching. (b) Photo images of experimental ITO transistor, Hall bar and TLM structures. 50 nm $SiO_2$ is used under source/drain pads to reduce the leakage current between gate and source/drain pads. (c) $R_{xy}$ *versus* B field in Hall measurement of 10 nm ITO with floating gate. Electron density of $1.7 \times 10^{20}$ /$cm^3$ and Hall mobility of 16.7 $cm^2$/V·s are extracted. (d) TLM measurement of 10 nm ITO with Ni contacts. Low contact resistance of 0.15 Ω·mm and low contact resistivity of $1.1 \times 10^{-7}$ Ω·$cm^2$ are achieved.

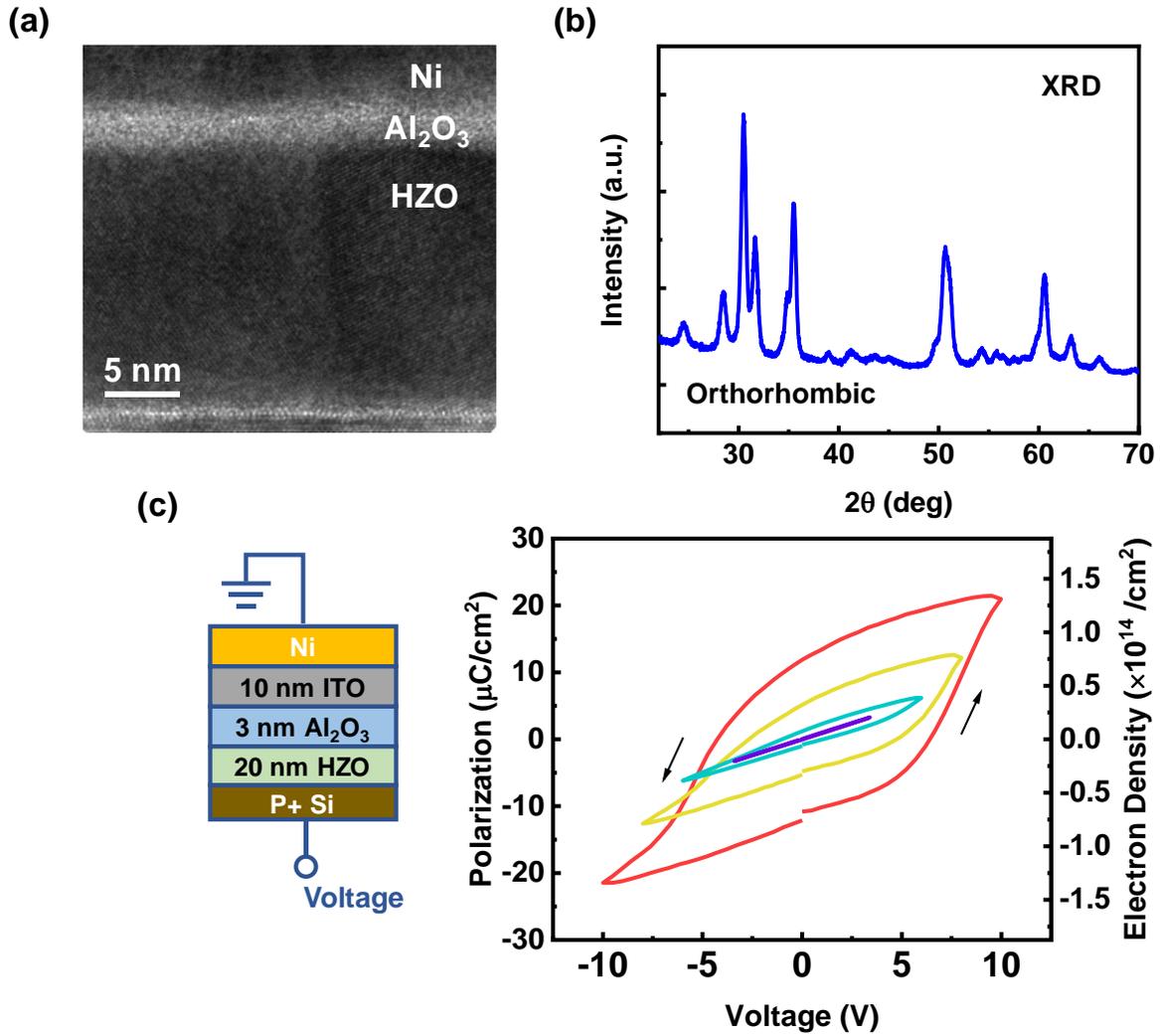

**Figure 2** (a) TEM image of Ni/Al$_2$O$_3$/HZO/Si stack, capturing amorphous Al$_2$O$_3$ and polycrystalline HZO. (b) XRD measurement of ferroelectric HZO, showing the HZO crystal contains orthorhombic phase. (c) P-V measurement of p+ Si/20 nm HZO/3 nm Al$_2$O$_3$/10 nm ITO stack, showing clear ferroelectric polarization switching. The maximum polarization is greater than 20 μC/cm$^2$, corresponding to a 2D electron density over $10^{14}$ /cm$^2$, indicating 10 nm ITO has a carrier density over $10^{14}$ /cm$^2$, which is consistent with Hall measurement in Fig. 1(c).

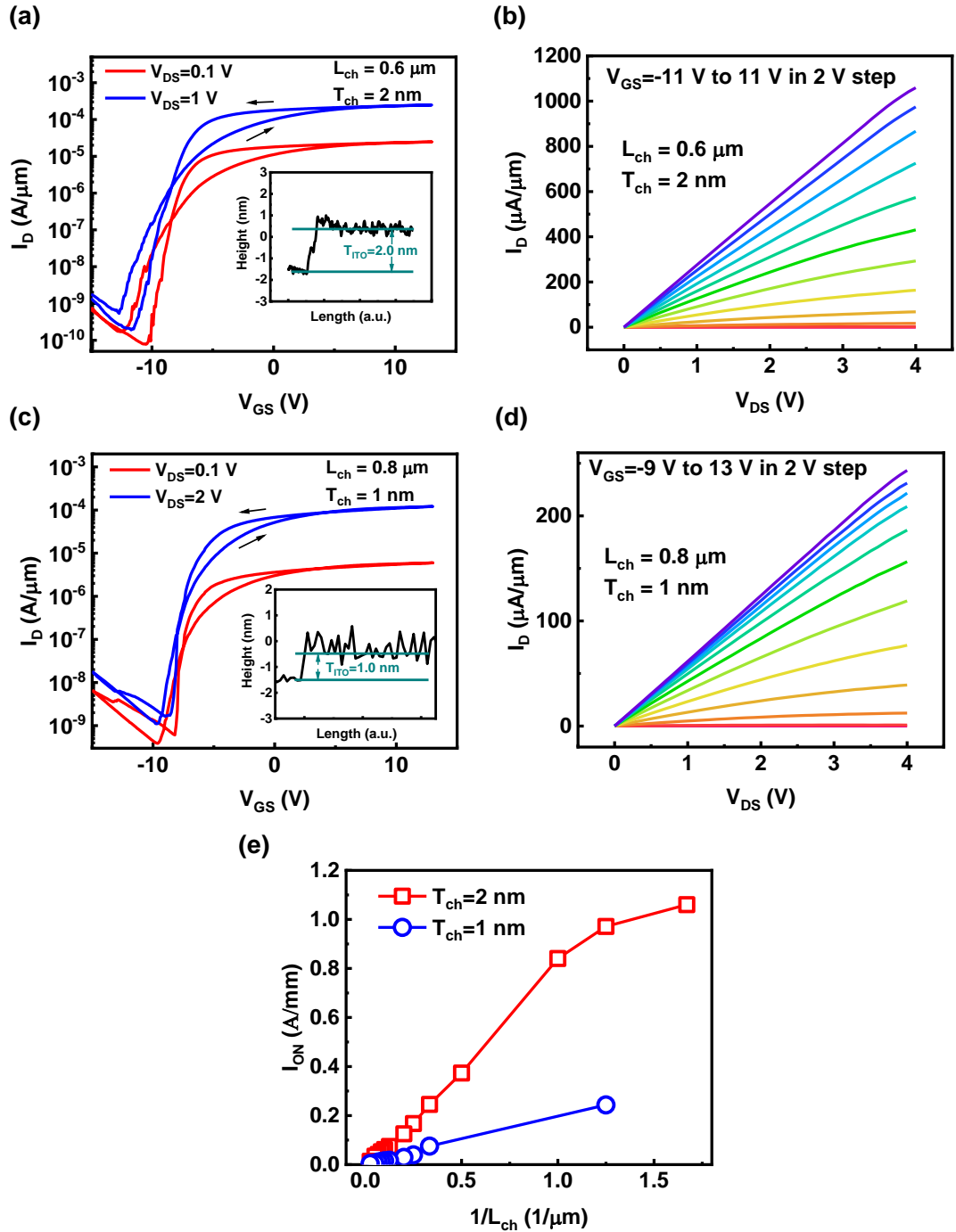

**Figure 3** (a) $I_D$-$V_{GS}$ and (b) $I_D$-$V_{DS}$ characteristics of an ITO transistor with channel length of 0.6 μm and channel thickness of 2 nm, exhibiting on-current of 1.06 A/mm and on/off ratio over 6 orders. (c) $I_D$-$V_{GS}$ and (d) $I_D$-$V_{DS}$ characteristics of an ITO transistor with channel length of 0.8 μm and channel thickness of 1 nm, exhibiting on-current of 0.243 A/mm. (e) $I_{ON}$ scaling metrics of ITO transistors with channel thicknesses of 1 nm and 2 nm at $V_{DS}$=4 V.

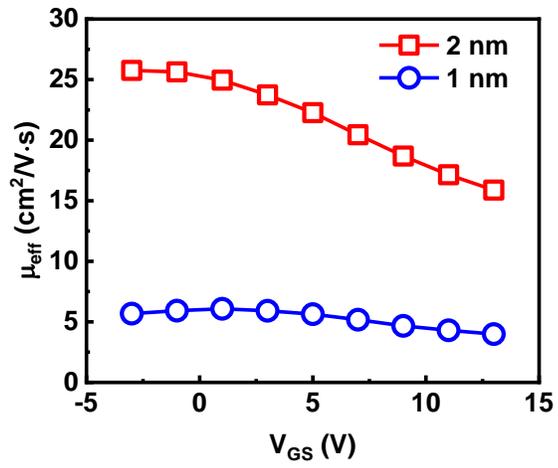

**Figure 4** Effective mobility *versus* gate voltage, extracted from output conductance of ITO transistors. $\mu_{eff}$ of 26.0 cm$^2$/V·s and 6.1 cm$^2$/V·s are achieved for 1-nm and 2-nm ITO.

TOC

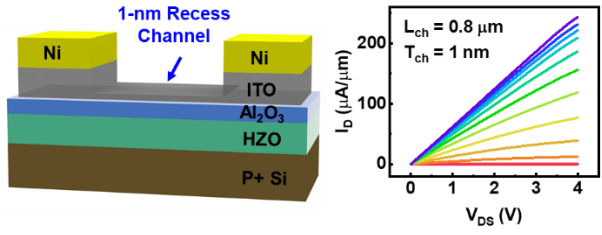

Supporting Information for:

# Indium-Tin-Oxide Transistors with One Nanometer Thick Channel and Ferroelectric Gating

Mengwei Si[1], Joseph Andler[2], Xiao Lyu[1], Chang Niu[1], Suman Datta[3], Rakesh Agrawal[4], Peide D. Ye[1,*]

[1]*School of Electrical and Computer Engineering and Birck Nanotechnology Center, Purdue University, West Lafayette, IN 47907, United States*

[2]*School of Materials Engineering, Purdue University, West Lafayette, IN 47907, United States*

[3]*Department of Electrical Engineering, University of Notre Dame, Notre Dame, IN 46556, United States*

[4]*Davidson School of Chemical Engineering, Purdue University, West Lafayette, IN 47907, United States*

* Address correspondence to: yep@purdue.edu (P.D.Y.)

## 1. Wet Etching of ITO Thin Film

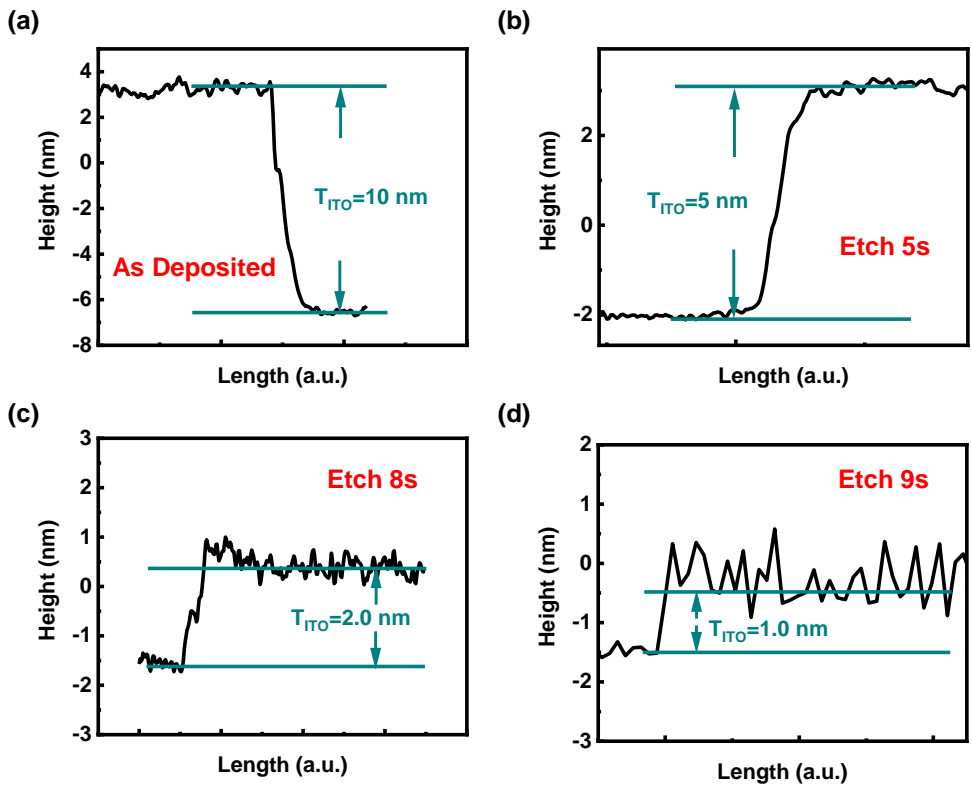

**Figure S1** AFM measurement on ITO films with different wet etching times: (a) as deposited, (b) 5s, (c) 8s, (d) 9s, by diluted HCl solution (3.4%).

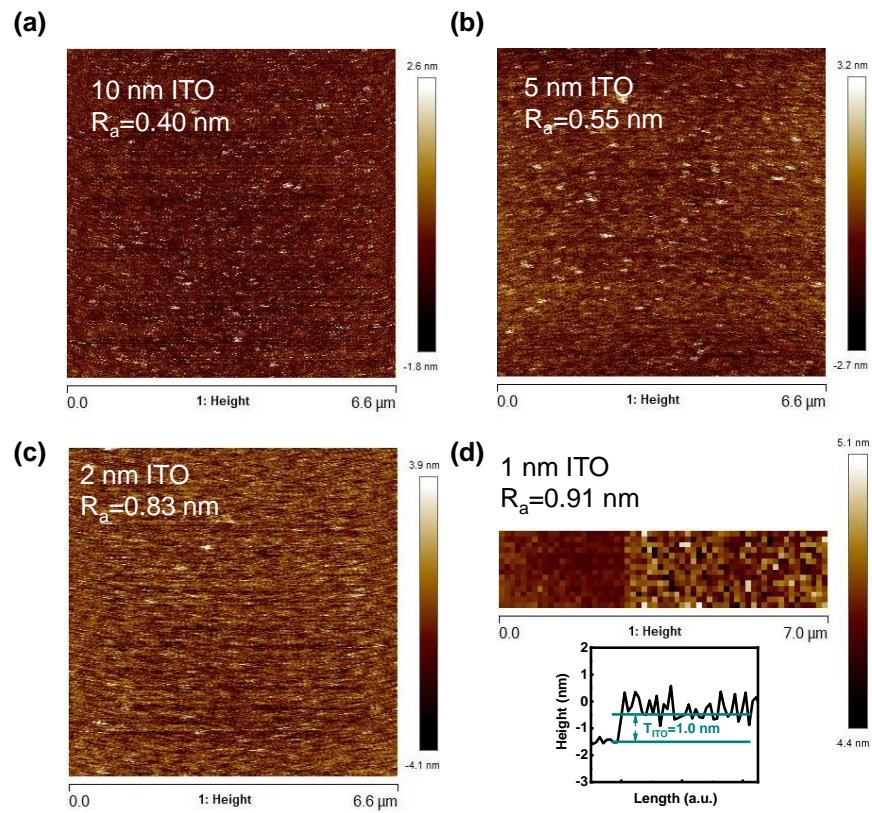

**Figure S2** Surface roughness by AFM measurement on ITO films with different thicknesses: (a) 10 nm as deposited, (b) 5 nm, (c) 2 nm, (d) 1 nm.

## 2. C-V Measurement of the Gate Stack

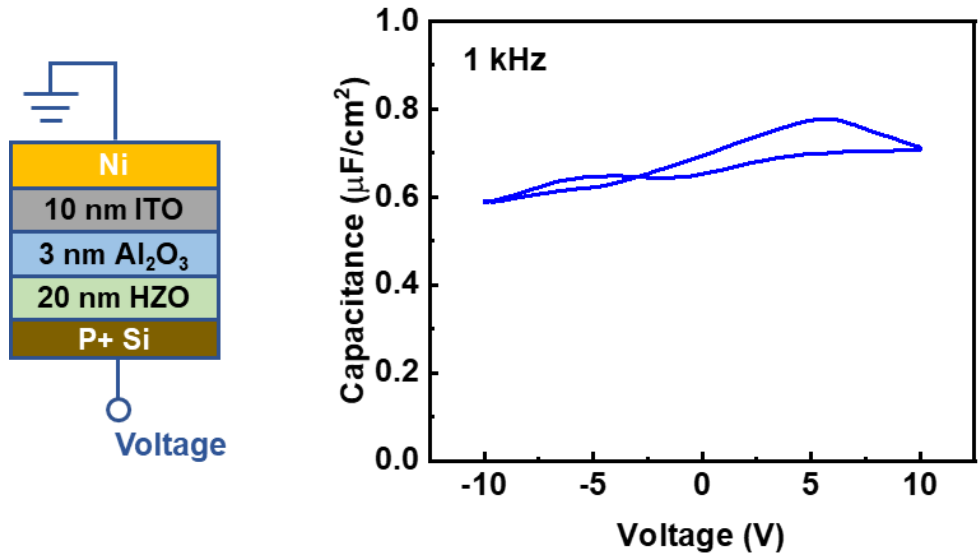

**Figure S3** C-V characteristics of p+ Si/20 nm HZO/3 nm Al$_2$O$_3$/10 nm ITO stack.

## 3. I-V Characteristics of ITO Transistors without Channel Recess

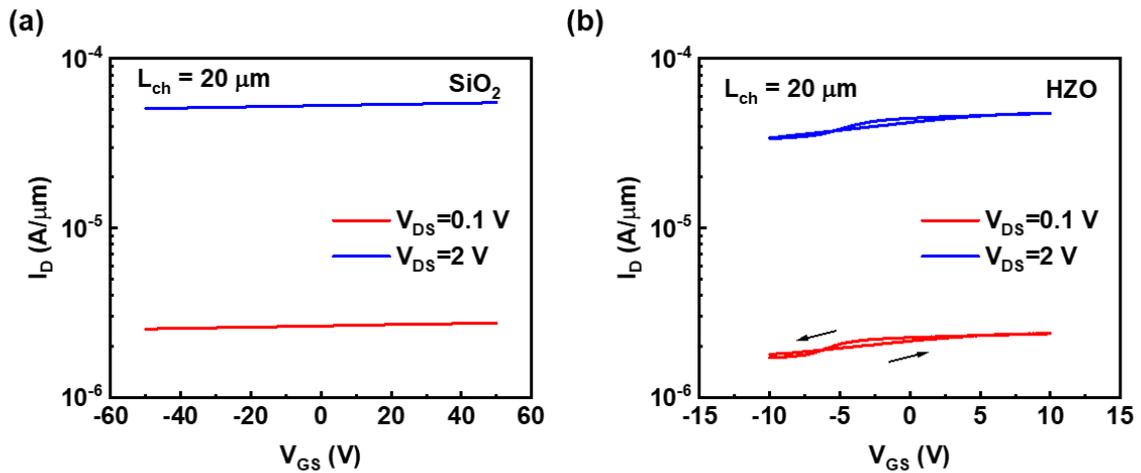

**Figure S4** $I_D$-$V_{GS}$ characteristics of ITO transistors with channel length of 20 μm and channel thickness of 10 nm. (a) 90 nm SiO$_2$ and (b) 20 nm HZO/3 nm Al$_2$O$_3$ are used as gate insulators, respectively.

## 4. I-V Characteristics of ITO Transistors with Dielectric Gate Insulator

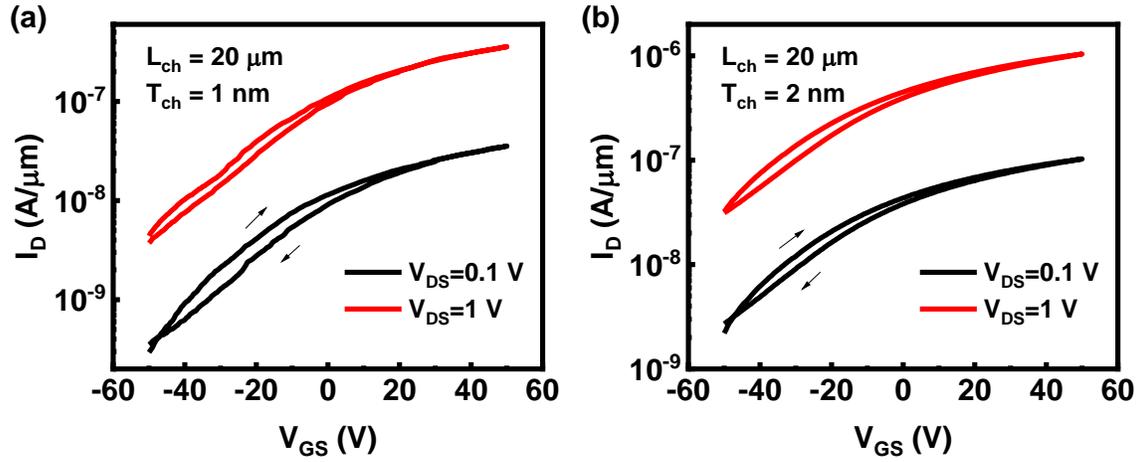

**Figure S5** $I_D$-$V_{GS}$ characteristics of ITO transistors with channel length of 20 μm and channel thickness of (a) 1 nm and (b) 2 nm. 90 nm $SiO_2$ is used as gate insulators.

$I_D$-$V_{GS}$ characteristics of ITO transistors with dielectric (DE) 90 nm $SiO_2$ as gate insulator and $L_{ch}$ of 20 μm and $T_{ch}$ of 1 nm and 2 nm are shown in Fig. S5(a) and S5(b). ITO transistors with FE HZO as gate insulator have much higher current density than ITO transistors with DE $SiO_2$ as gate insulator, indicating the polarization charges in FE HZO enhance the carrier density in ITO channel.

## 5. Minor Loops in ITO Transistor with Ferroelectric Gating

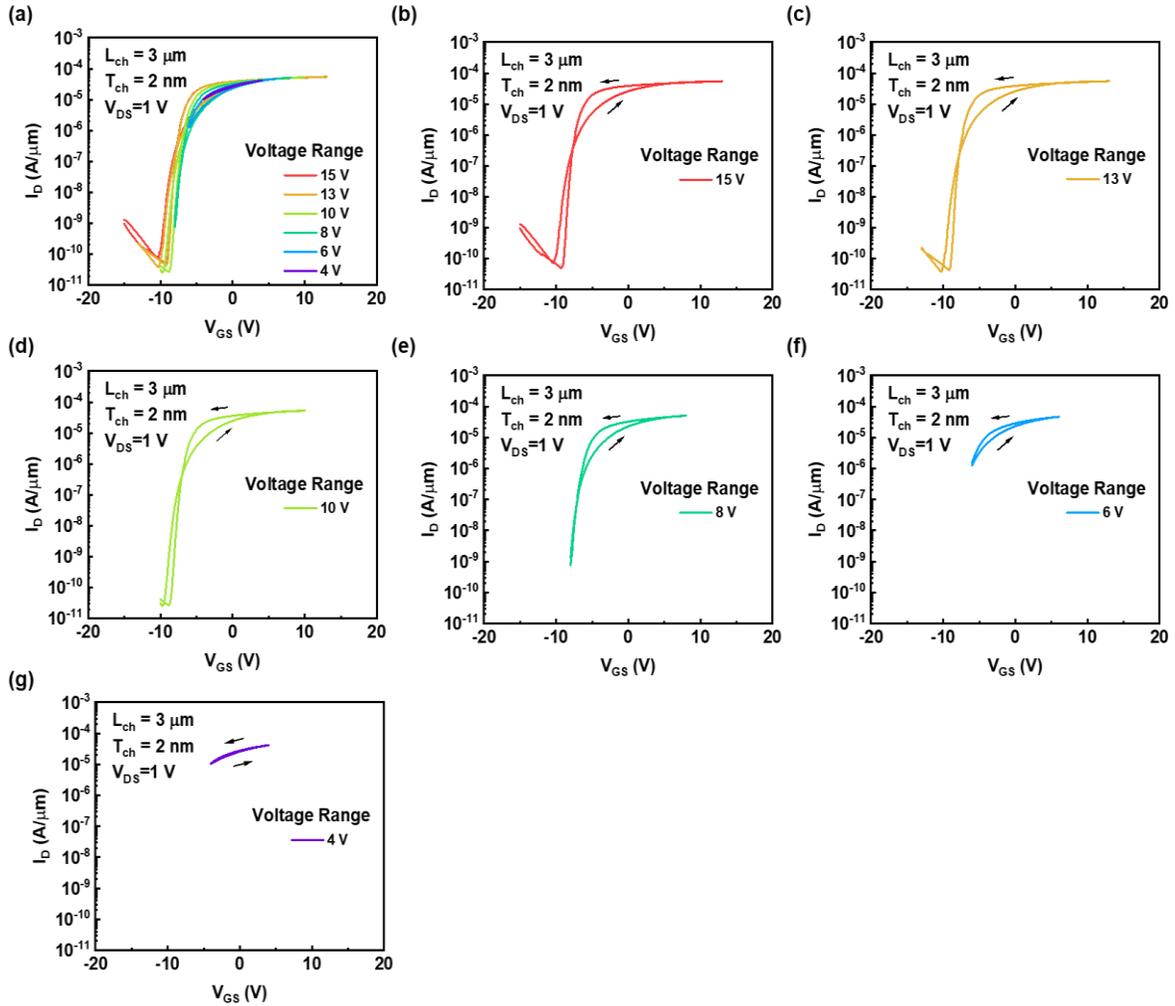

**Figure S6** $I_D$-$V_{GS}$ characteristics of an ITO transistor with channel length of 3 μm and channel thickness of 2 nm,

Fig. S6(a) shows the $I_D$-$V_{GS}$ characteristics of an ITO transistor with $L_{ch}$ of 3 μm and $T_{ch}$ of 2 nm at $V_{DS}$=1 V and at different voltage sweep ranges (from 15 V down to 4 V). Fig. S6(b)-(g) are the individual double sweep $I_D$-$V_{GS}$ curves at different voltage sweep range. As we can see, the counterclockwise hysteresis loop becomes smaller at reduced voltage sweep range, indicating less polarizaiton charge. The device doesn't exhibit memory window as large as ferroelectric field-effect transistors made of Si or Ge,[1,2] suggesting ferroelectric polarization switching with a wideband gap semiconductor channel may be fundamentally different.

## 6. Channel Length Scaling of ITO Transistor

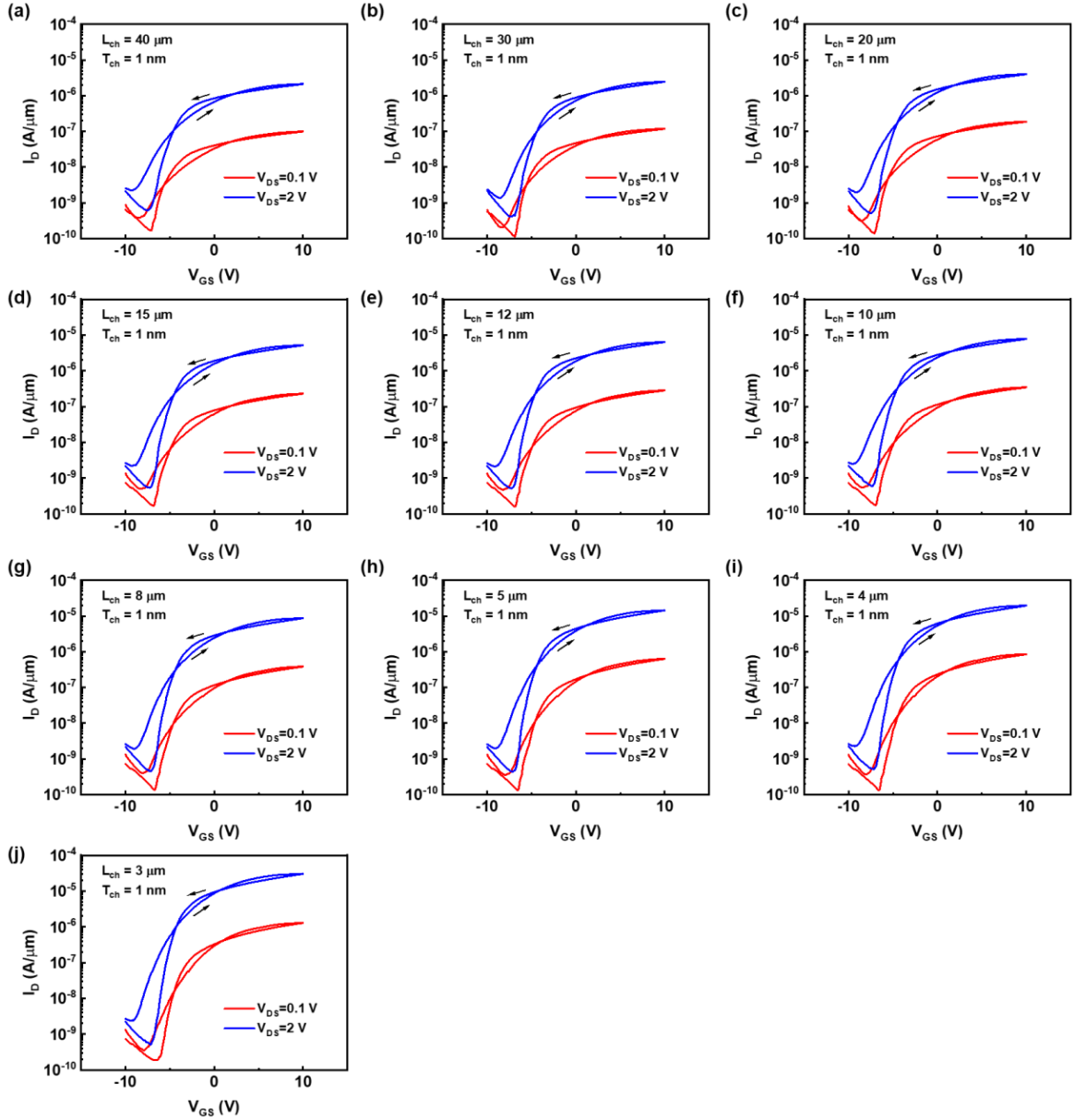

**Figure S7** $I_D$-$V_{GS}$ characteristics of ITO transistors with channel thickness of 1 nm and channel length from 40 μm down to 3 μm.

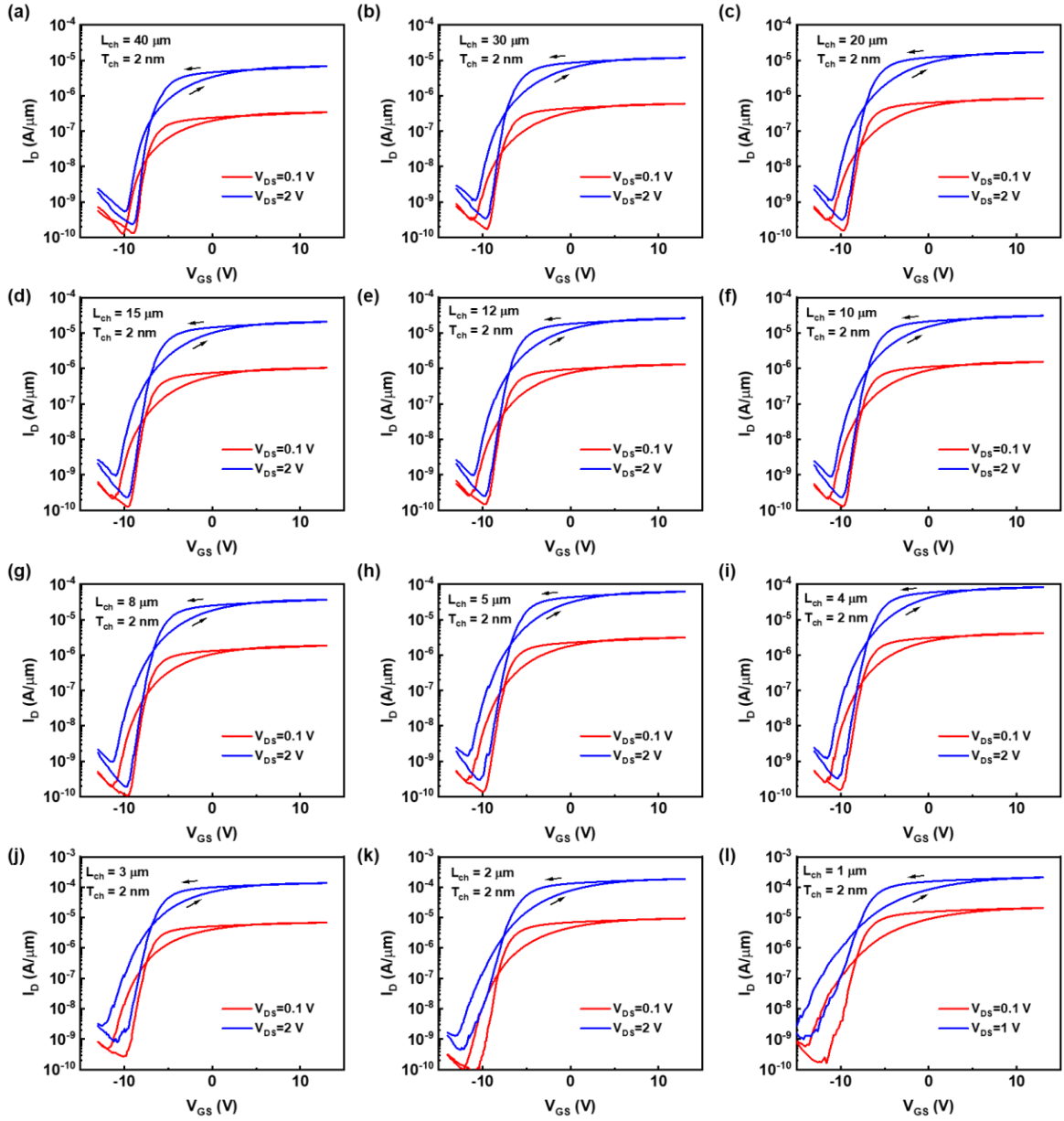

**Figure S8** $I_D$-$V_{GS}$ characteristics of ITO transistors with channel thickness of 2 nm and channel length from 40 µm down to 1 µm.

## 7. Series Resistance in ITO transistors

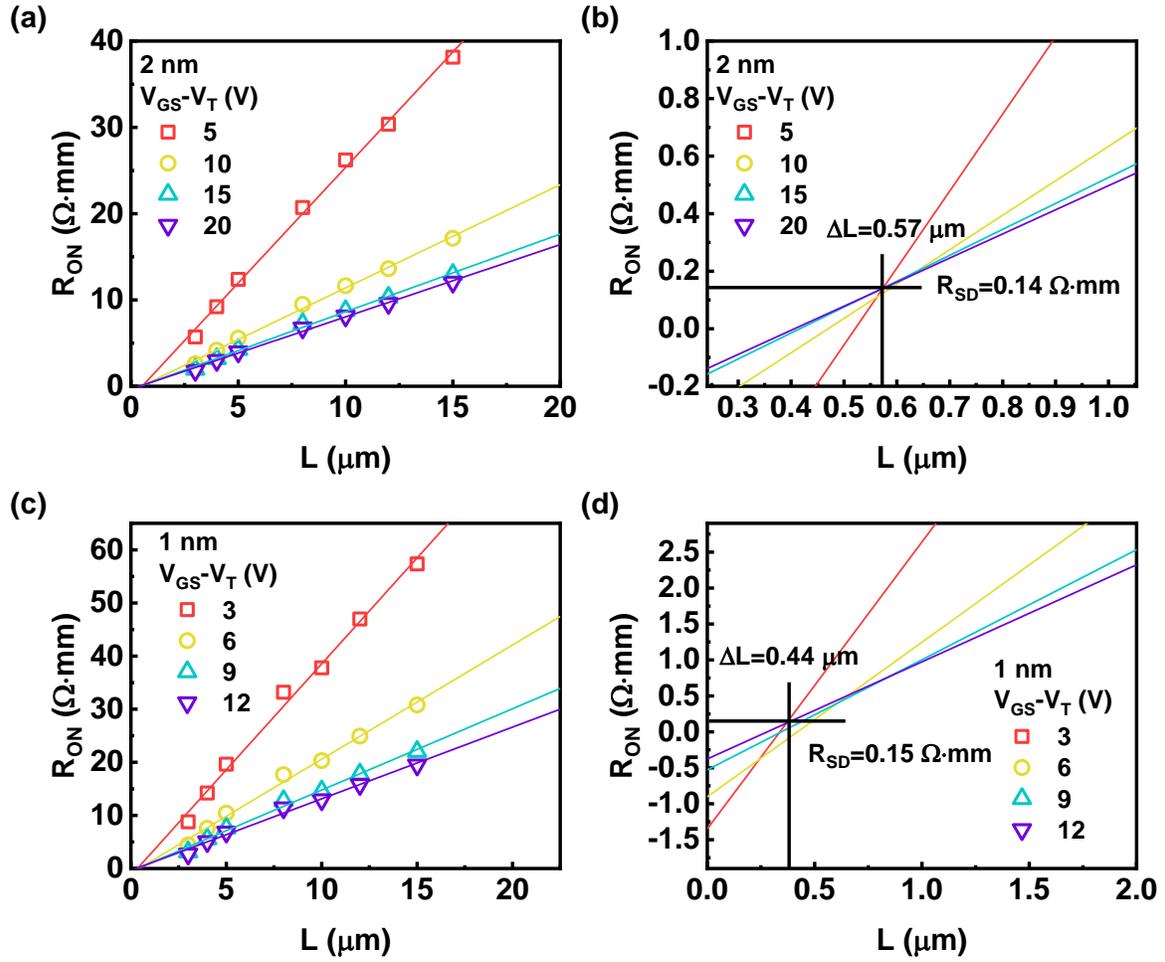

**Figure S9** (a) and (c) on-resistance *versus* channel length for ITO transistors with $T_{ch}$ of 2 nm and 1 nm at different $V_{GS}$-$V_T$. (b) and (d) Zoomed on-resistance *versus* channel length of (a) and (c). $R_{SD}$ of ITO transistors with 2-nm ITO is extracted to be 0.14 Ω·mm and $R_{SD}$ of ITO transistors with 1-nm ITO is extracted to be 0.15 Ω·mm.

## 8. Carrier Density of ITO transistors

$\mu_{eff}$ of 26.0 cm$^2$/V·s and 6.1 cm$^2$/V·s are achieved for 2-nm and 1-nm thick ITO, as shown in Fig. 4. Carrier density can be estimated according to $I_D=n_{2D}q\mu E$, where $n_{2D}$ is the 2D carrier density, q is the elementary charge, µ is the mobility, E is the source to drain electric field. According to this equation, the carrier density in 1-nm and 2-nm ITO channel can be calculated as shown in Fig. S10. Carrier density of 1-nm and 2-nm ITO are similar, with maximum $n_{2D}$ about $0.8\times10^{14}$ /cm$^2$. Therefore, the current density difference between 1-nm and 2-nm ITO comes from the mobility difference. The reduction of drain current density in devices with 1-nm ITO channel is the result of mobility degradation from surface scattering. For 2-nm ITO transistor with 90 nm SiO$_2$ as gate insulator (Fig. S5(b)), according to $\mu_{eff}$=26.0 cm$^2$/V·s, a maximum $n_{2D}$ of $4.9\times10^{12}$ /cm$^2$ is obtained. For 1-nm ITO transistor with 90 nm SiO$_2$ as gate insulator (Fig. S5(a)), according to $\mu_{eff}$=6.1 cm$^2$/V·s, a maximum $n_{2D}$ of $7.2\times10^{12}$ /cm$^2$ is obtained. $n_{2D}$ is more than one order of magnitude higher in ITO transistor with FE gating ($n_{2D} > 8\times10^{13}$ /cm$^2$). Such difference cannot be the result of different EOT. For example, for 50 V on 90 nm SiO$_2$, the voltage/EOT is about 0.6 V/nm; for 13 V on 20 nm HZO/3 nm Al$_2$O$_3$, the voltage/EOT is about 2.4 V/nm. The difference in displacement field is only 4 times. Therefore, only EOT difference itself cannot lead to the enhancement demonstrated in this work. Considering the low current density in ITO transistors with SiO$_2$ as gate insulator, as shown in Fig. S5(a) and S5(b), the high carrier density in ITO transistor with FE HZO as gate insulator comes from the enhancement by FE polarization.

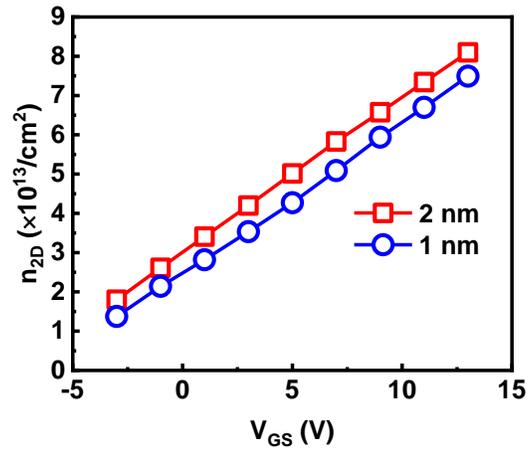

**Figure S10** Carrier density *versus* gate voltage for 1-nm and 2-nm ITO.

## 9. Device Variations

The ITO deposition was done by sputtering so that this technology can be used for large-area fabrication process. Fig. S11 shows $I_D$-$V_{GS}$ characteristics of 13 ITO transistors with channel thickness of 2 nm and channel length of 3 μm, showing similar switching characteristics. The device-to-device variation comes from variation in wet etching process. Such variation can be further improved by introducing dry etching or re-growth source/drain process.

The off-state current in this work comes from the gate leakage current through FE HZO, as shown in Fig. S12. $I_G$ and $I_D$ at off-state are very similar. Therefore, the off-state current variation in this work originates from the gate leakage current variation.

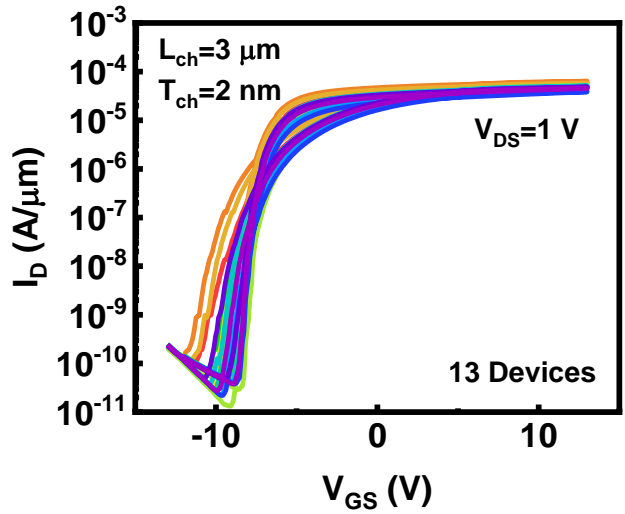

**Figure S11** $I_D$-$V_{GS}$ characteristics of 13 ITO transistors with channel thickness of 2 nm and channel length of 3 μm.

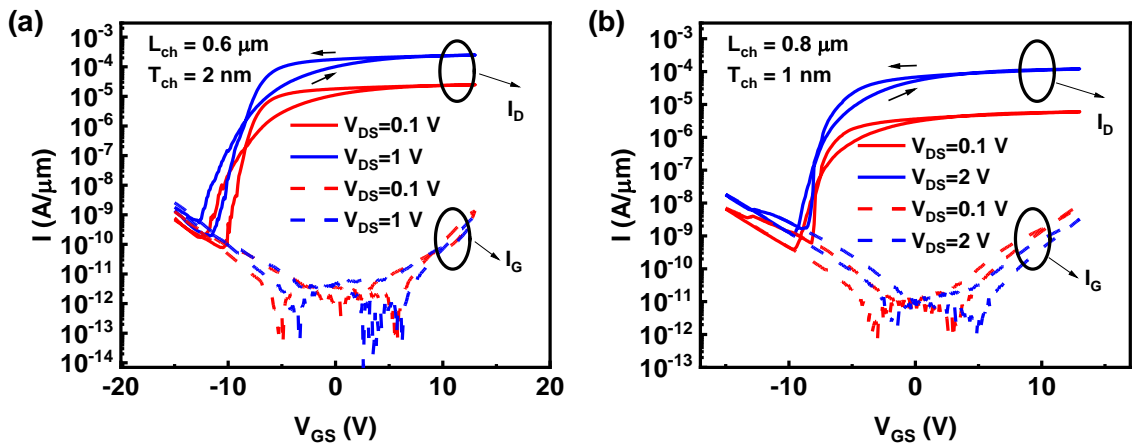

**Figure S12** $I_D$-$V_{GS}$ and $I_G$-$V_{GS}$ characteristics of (a) an ITO transistor with channel thickness of 2 nm and channel length of 0.6 μm and (b) an ITO transistor with channel thickness of 1 nm and channel length of 0.8 μm.